\documentclass[runningheads]{llncs}
\usepackage[T1]{fontenc}
\usepackage{graphicx}
\usepackage{epstopdf}
\usepackage[misc]{ifsym}
\usepackage{multirow}
\usepackage[marginal]{footmisc}
\usepackage{amsmath}
\usepackage{amssymb}
\usepackage[tight,footnotesize]{subfigure}
\usepackage{color,xcolor}
\usepackage{wrapfig}
\usepackage{caption}

\begin{document}
%
\title{Fuzzy Attention-based Border Rendering Network for Lung Organ Segmentation}
\titlerunning{Fuzzy Attention-based Border Rendering Network for Lung Organ Segmentation}
%
\author{
Sheng Zhang\inst{1}* \and
Yang Nan\inst{1}* \and
Yingying Fang\inst{1}* \and
Shiyi Wang\inst{1} \and
Xiaodan Xing\inst{1} \and
Zhifan Gao\inst{2} \and
Guang Yang\inst{1}\Letter
}
\authorrunning{Sheng et al.}
%
\vspace{-3.0em}
\institute{
$^1$Imperial College London, London, UK; $^2$Sun Yat-sen University, Shenzhen, China\\
\email{g.yang@imperial.ac.uk
}
}
%
\maketitle              
\footnote{*Equal contribution.}
\vspace{-3.0em}
\begin{abstract}
Automatic lung organ segmentation on CT images is crucial for lung disease diagnosis. However, the unlimited voxel values and class imbalance of lung organs can lead to false-negative/positive and leakage issues in advanced methods. Additionally, some slender lung organs are easily lost during the \textit{recycled} down/up-sample procedure, e.g., bronchioles \& arterioles, causing severe discontinuity issue. Inspired by these, this paper introduces an effective lung organ segmentation method called Fuzzy Attention-based Border Rendering (FABR) network. Since fuzzy logic can handle the uncertainty in feature extraction, hence the fusion of deep networks and fuzzy sets should be a viable solution for better performance. Meanwhile, unlike prior top-tier methods that operate on all regular dense points, our FABR depicts lung organ regions as cube-trees, focusing only on \textit{recycle}-sampled border vulnerable points, rendering the severely discontinuous, false-negative/positive organ regions with a novel Global-Local Cube-tree Fusion (GLCF) module. All experimental results, on four challenging datasets of airway \& artery, demonstrate that our method can achieve the favorable performance significantly.\\
\vspace{-1.5em}
\end{abstract}

\vspace{-2.2em}
\section{Introduction}
\vspace{-1.0em}
Automatic lung organ segmentation is one of the challenging tasks in the field of medical image analysis \cite{ronneberger2015u,lin2021seg4reg+,nan2024hunting}. Recently, this task has been extended to variously realistic applications, e.g., robotic surgery \cite{gao2021future}, lung disease diagnosis \& prognosis \cite{tsay2021lower,fang2024dynamic}. To achieve a superb segmentation performance, it is vital to learn a group of abundant and salient descriptions of lung image feature.
\begin{figure}[!htb]
\vspace{-1.0em}
\centering
\includegraphics[width=0.8\textwidth]{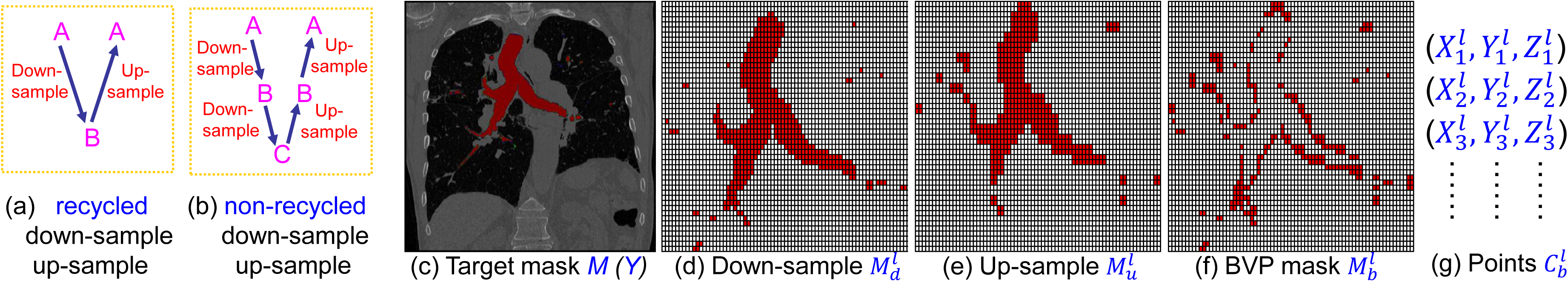}
\vspace{-0.5em}
\caption{The elaboration of \textbf{B}order \textbf{V}ulnerable \textbf{P}oints (BVP) caused by \textbf{recycled} down-sample and up-sample in the encoder-decoder backbone. Downsampling (c) gets (d), upsampling (d) gets (e), then (f) is the absolute difference of (c) \& (e). In the test phase, (c) is binarized coarse prediction.}\label{fig:bvp}
\vspace{-1.0em}
\end{figure}
However, current state-of-the-art methods of lung organ segmentation still face several challenges and aspects for improvement.
Firstly, the unlimited voxel values, multi-site imaging discrepancy and class imbalance in lung organ images can lead to false-negative and leakage issues in prior segmentation methods, which badly influences the critical early diagnosis of imperceptible lung diseases, e.g., lung fibrosis, nodule and hypertension, etc. Secondly, the presence of numerous slender branches, e.g., bronchioles and arterioles, which are easily lost during the \textit{recycled} down/up-sampling procedure in Fig. \ref{fig:bvp}, can result in discontinuity, detail loss, and coarse mask predictions. Thirdly, most CNN-based medical segmentation methods treat all points equally during the mask rendering stage, overlooking the vulnerability of border points in Fig. \ref{fig:bvp} (f) and the importance of explicit border modeling. Lastly, while Vision Transformer (ViT) has shown promise in computer vision tasks \cite{dosovitskiy2020image,hatamizadeh2022unetr}, its quadratic operation complexity limits its application in 3D high-resolution CT images due to hardware constraints. Meanwhile, most specific datasets for medical image analysis are small and scarce due to laborious manual annotation and privacy protection, which badly restricts the potential of transformer-based top-tier methods.

To address these limitations in this paper, we propose an effective lung organ segmentation method called FABR. Unlike prior approaches, the method FABR fuses fuzzy sets and deep network to diminish the uncertainty in feature representations, decouples and depicts medical image regions as cube-trees, specifically targeting the border vulnerable points illustrated in Fig. \ref{fig:arch1}. To address the challenges of severe discontinuity and false-negative/positive bronchioles and arterioles, we propose one innovative module of global-local cube-tree fusion, which fuses the learnable global embedding and local lung organ features.

In summary, our main contributions are three-folds: (1) We seamlessly integrate efficient fuzzy attention theory and transformer-like expansion/compression convolutional network to diminish the uncertainty of lung organ feature representations; (2) We present an innovative global-local cube-tree fusion module, which explicitly models the border vulnerable points yielded by recycled down/up-sample for accurate lung organ segmentation; (3) We do extensive experiments on four challenging datasets to prove the efficacy of our method.

\vspace{-0.5em}
\section{Methodology}
\begin{figure}[!htb]
\vspace{-1.5em}
\centering
\includegraphics[width=0.9\textwidth]{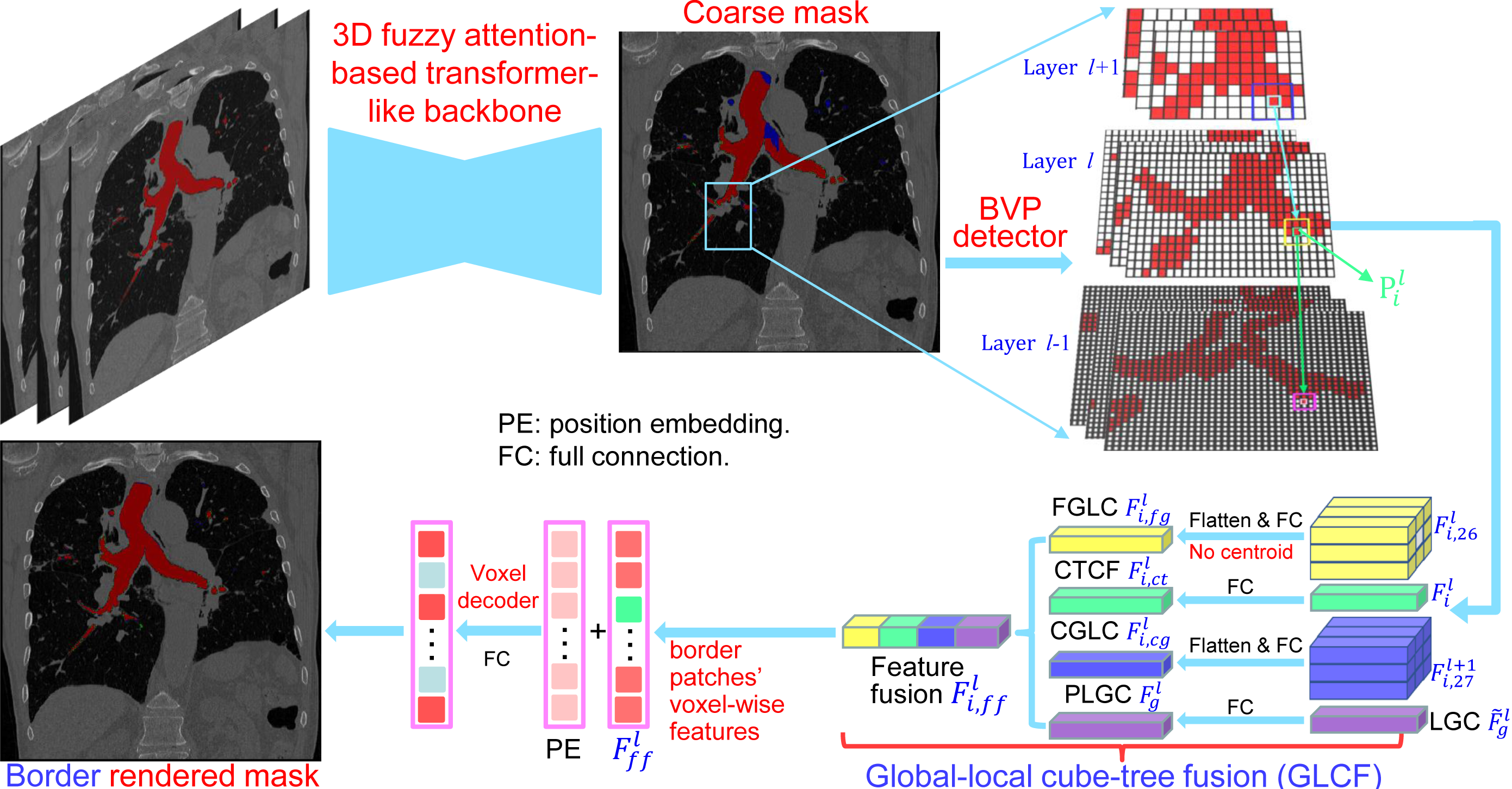}
\vspace{-0.5em}
\caption{The overview of our method FABR. FGLC: fine grain local context; CTCF: cube-tree centroid feature; CGLC: coarse grain local context; PLGC: projected learnable global context. BVP detector is shown in Fig. \ref{fig:bvp}. Noting the matched relationship between top-right boxes' and bottom-right bars' colors.}\label{fig:arch1}
\vspace{-1.5em}
\end{figure}

\begin{figure}[!htb]
\vspace{-1.5em}
\centering
\includegraphics[width=0.9\textwidth]{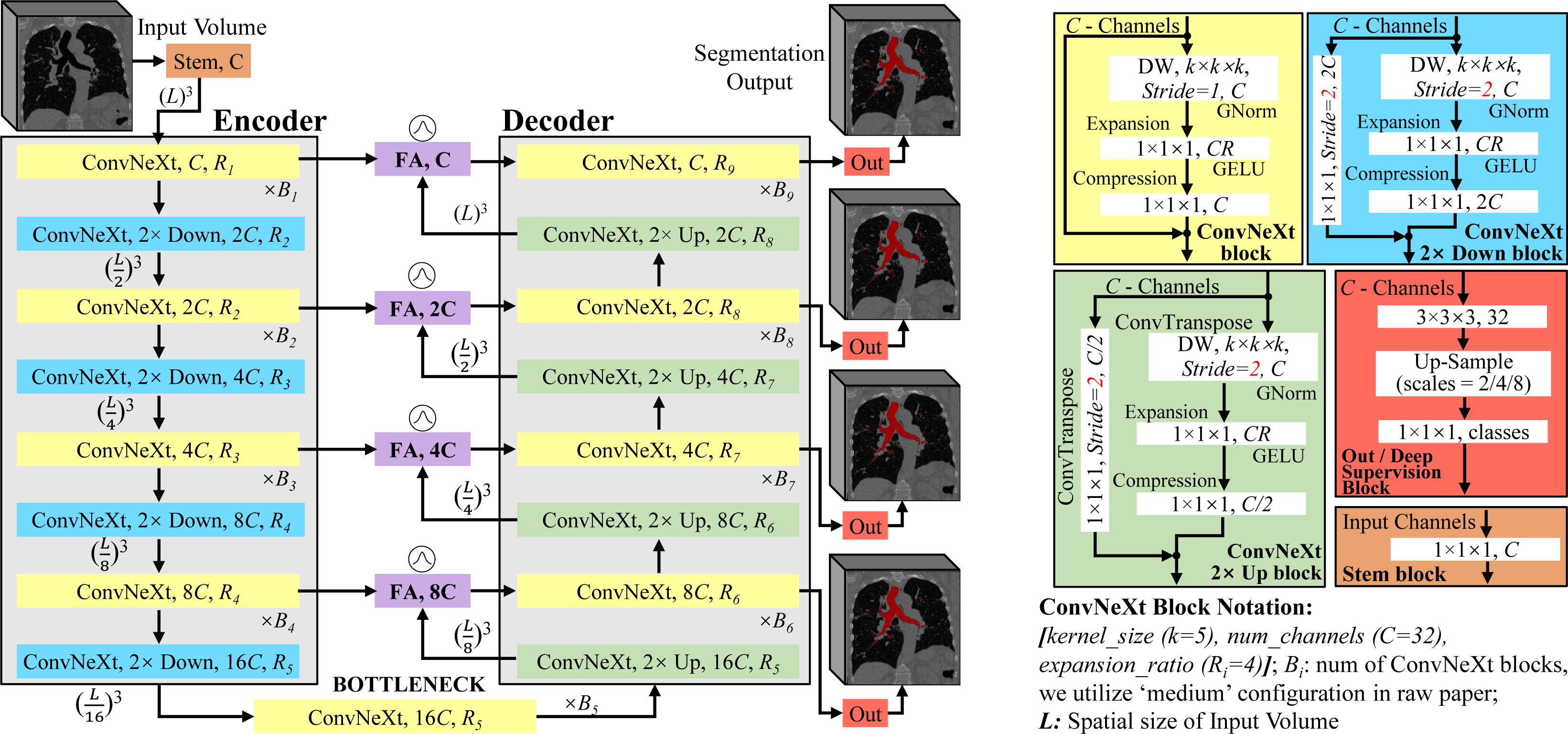}
\vspace{-0.5em}
\caption{Our FA-based transformer-like backbone design and coarse mask generation. FA: fuzzy attention module detailed in subsection \ref{sec:fa}. DW: depth-wise convolution. }\label{fig:backbone1}
\vspace{-1.5em}
\end{figure}

\vspace{-0.5em}
The overview of our method FABR is detailed in Fig. \ref{fig:arch1}. It mainly includes two modules, i.e., fuzzy attention-based transformer-like 3D U-shaped backbone and \textbf{G}lobal-\textbf{L}ocal \textbf{C}ube-tree \textbf{F}usion (GLCF) module.
The fuzzy attention-based transformer-like backbone is inspired by the well-known ConvNeXt \cite{woo2023convnext} and detailed in Fig. \ref{fig:backbone1}, which includes a preliminary stem, sequential transformer-like regular/down/up-sample convolution blocks, a bottleneck and four efficient fuzzy attention modules, where each convolution block is constructed by applying a large kernel of $5\times5\times5$ 3D separable depth-wise convolution/deconvolution, group-normalization, transformer-like architecture (i.e., embedding 4$\times$ expansion/compression $1\times1\times1$ convolution layers like FFN module of transformer in our Fuzzy attention module) and GELU activation layer. The corresponding layers of the same scale between the encoder and decoder are linked by the efficient fuzzy attention layer. Besides, each-scale stage of the decoder is added by the $1\times1\times1$ 3D convolution and activation layers to predict the preliminary coarse masks of lung organ segmentation. Then, unlike the prior top-tier methods that operate on all regular dense points of the coarse masks to render the raw prediction, the proposed GLCF module decouples and depicts the medical image regions as cube-trees, which only focuses on the \textit{recycle}-sampled BVP, and renders the severe discontinuity as well as false-negative/positive bronchioles or arterioles. We now elaborate the insights within the proposed method FABR for each innovative module in the following subsections.

\vspace{-0.5em}
\subsection{Fuzzy Attention-based Transformer-like Backbone}\label{sec:fa}
\vspace{-0.5em}
One of the key challenges to design a robust lung organ segmentation module lies in the inherent uncertainty from the organ annotations and voxel values, e.g., bronchioles and arterioles. Various efforts have been done to enhance the network to focus on pertinent regions. Notably, Attention U-Net \cite{oktay2018attention} introduces an attention gate to bolster accuracy by suppressing feature activations in irrelevant regions. However, we deem that the non-channel specifics of current attention map assign the same ``attention'' coefficient to all feature points along the channel dimension. Specifically, given a feature map \(F\in{\mathbb{R}^{C\times{H}\times{W}\times{D}}}\), the extant attention map is built as \(\alpha\in{\mathbb{R}^{H\times{W}\times{D}}}\), while all features along the channel wise C share the same ``importance''. This mechanism is unreliable since the features in different channels are extracted by different convolution kernels; therefore, we advocate the attention map to be channel-specific.

\begin{wrapfigure}{r}{0.6\textwidth}
\vspace{-1.5em}
\centering
\includegraphics[width=0.6\textwidth]{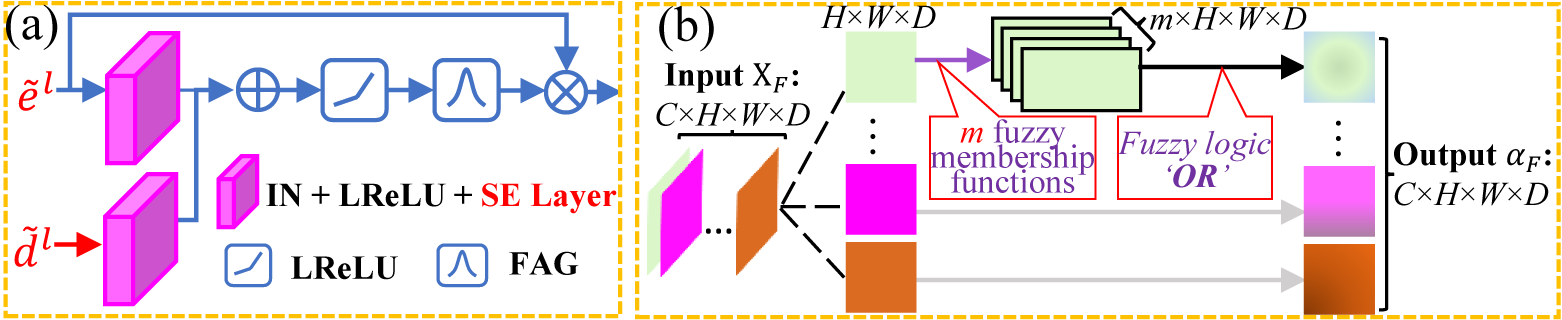}
\vspace{-1.5em}
\caption{The details of (a) our efficient fuzzy attention module and (b) fuzzy attention gate (FAG) in the subfigure (a). Zooming in for a better view.}\label{fig:fal}
\vspace{-1.5em}
\end{wrapfigure}

Meanwhile, numerous studies have proved the efficacy of both fuzzy logic and neural networks in data representation \cite{nan2023fuzzy}. Broadly speaking, neural networks strive to diminish noise in original data to extract meaningful feature representations, while fuzzy logic can derive fuzzy representations, mitigating the original data uncertainty. Hence, we fuse fuzzy logic with attention mechanism by utilizing trainable Gaussian membership functions (GMFs). This fusion serves to enhance the segmentation network's ability to focus on pertinent regions, concurrently diminishing uncertainty and variations in data representations.

As shown in Fig. \ref{fig:fal}(a), the proposed efficient fuzzy attention module is adopted within the skip connection, taking both feature maps \{\(\tilde{e}^l\), \(\tilde{d}^l\)\} from the \textit{l-th} encoder and decoder layers as inputs, which are directly yielded by the transformer-like 4$\times$ expansion/compression layers in ConvNeXt \cite{woo2023convnext} backbone, followed by an instance normalization and a Leaky-ReLU layers for feature reconstitution. Then, two very lightweight squeeze-excitation (SE) layers \cite{hu2020squeeze} are employed to further boost the channel-specificity. Next, a voxel-wise adding operation is adopted to fuse the information, followed by a Leaky-ReLU. Eventually, the feature representations are fed into the FAG to generate a voxel-wise attention map, shown in Fig. \ref{fig:fal}(b). Assume \(X\) $\in$ \(R^{C\times{H}\times{W}\times{D}}\) (regardless of batch size) as the input of FAG. Due to the smoothness and concise notation of GMFs, learnable GMFs are proposed to specify the deep fuzzy sets. Each feature map (with size \(H\times W\times D\)) is filtered by \(m\) GMFs with the trainable centre \(\mu_{i,j}\) and spread \(\sigma_{i,j}\)
\vspace{-0.5em}
\begin{equation}\label{eq:fz_ga}
    f_{i,j}(X,\mu,\sigma)=e^{(-(X_j-\mu_{i,j})^2)/(2\sigma_{i,j}^2)},
    \vspace{-0.5em}
\end{equation}
where $i\in\{1,\ldots,m\},j\in\{1,\ldots,C\}$. Our goal is to use the $m$ membership function to learn the ``importance'' of target fuzzy feature representations. Given the trade-off of model efficiency \& efficacy, $m = 4$ GMFs are used. Thus, we assume that the information can be better preserved by applying the aggregation operator ``OR'' while suppressing irrelevant features. Given fuzzy sets \(\Tilde{A}\) and \(\Tilde{B}\), the operator ``OR'' is denoted as Equ. \ref{eq:fz_or}(a).
\vspace{-0.5em}
\begin{equation}
    f_{\Tilde{A}\cup\Tilde{B}}(y)=f_{\Tilde{A}}(y)\vee f_{\Tilde{B}}(y),~ \forall{y\in{U}}, (a);\ \ f_{\Tilde{A}\cup\Tilde{B}}(y)=\max(f_{\Tilde{A}}(y), f_{\Tilde{B}}(y)), (b)
    \label{eq:fz_or}
    \vspace{-0.5em}
\end{equation}
where U is the universe of information and $y$ is the element of U. To make the operator ``OR'' derivative, we modified it as Equ. \ref{eq:fz_or}(b).
Then, the fuzzy degree \(f_j(X,\mu,\sigma)\in{\Theta^{H\times{W}\times{D}}}, \Theta\in{[0,1]}\)
of the \(j\)-th channel can be obtained based on Equ. \eqref{eq:fz_ga} and Equ. \eqref{eq:fz_or} as
\vspace{-0.5em}
\begin{equation}
f_{j}(X,\mu,\sigma)=\bigvee_{i=1}^{m} {e^{\frac{-(X_j-\mu_{i,j})^2}{2\sigma_{i,j}^2}}}=\max(e^{\frac{-(X_j-\mu_{i,j})^2}{2\sigma_{i,j}^2}}),
\vspace{-0.5em}
\end{equation}
where $\bigvee$ indicates the union operation. Finally, the output tensor of proposed FAG has the same shape as input $X$, providing a voxel-wise attention map $\alpha^F$.

\vspace{-0.5em}
\subsection{Global-Local Cube-tree Fusion}
\vspace{-0.5em}
To the best of our knowledge, most mask render-based two-stage semantic segmentation methods \cite{isensee2021nnu,yang2022st++} operate equally on all dense points of the coarse masks to improve the final performance, which is unnecessary to focus much on the already correctly predicted points.
As shown in Fig. \ref{fig:bvp} and according to our statistical error analysis, most very vulnerable points occur on the object border due to the information loss caused by down-sample operation in the encoding process, especially for the innumerable bronchioles or arterioles in the tree-like structures. Thus, we only focus on the border vulnerable points and propose the novel global-local cube-tree fusion module. Specifically, (1) we ``\textbf{recycle}'' the down-sample and up-sample operations to produce masks $M_d^l$ and $M_u^l$, and evaluate the absolute difference $M_b^l$ of them in Fig. \ref{fig:bvp} to get the border vulnerable points $C_b^l$ for the $l$-th layer; (2) as shown in the top-right side of Fig. \ref{fig:arch1}, we build the cube-tree of the $i$-th point $P_{i}^l$ $\in$ $C_b^l$ by extracting the local contextual features $\{F_{i,26}^l,F_{i,27}^{l+1}\}$ of \{26, 27\}-neighbors of the \{$l$, $l$+1\}-th layers respectively, which are defined as the $3\times 3 \times 3$ cube without and with centroid. For the last layer, it is of note that we extract the 27-neighbors' local contextual features $F_{i,27}^{l-1}$ in the adjacent layer $l$-1; (3) we flatten features $\{F_{i,26}^l,F_{i,27}^{l+1}\}$ in the spatial dimension and project them as well as centroid feature $F_{i}^l$ into three vectors $\{F_{i,fg}^l,F_{i,cg}^l,F_{i,ct}^l\}$, which are separately related to the fine grain, coarse grain local context information and cube-tree centroid feature; (4) global airway or artery features from the distribution of the whole dataset is also very important, hence, we introduce the learnable global features $\tilde{F}_{g}^l \in R^d$ to yield the projected global features $F_{g}^l$, where $d \in \{32, 64, 128, 256\}$ is the embedding dimension; (5) we fuse the four features into $F_{i,ff}^l$ as follows:
\vspace{-0.6em}
\begin{equation}\label{equ:ff_mgf0}
\centering
F_{i,ff}^l =\lambda_1 F_{i,cg}^l + \lambda_2 F_{i,ct}^l + \lambda_3 F_{i,fg}^l + \lambda_4 F_{g}^l,
\vspace{-0.55em}
\end{equation}
where $\lambda_1 \sim \lambda_4$ $\in$ [0, 1] are the learnable coefficients to balance the importance of each feature; (6) we lastly add the feature $F_{i,ff}^l$ to the relative position embedding features $F_{i,pe}^l$ $\in$ \(R^{C_1\times{H}\times{W}\times{D}}\) (retaining the topology information for inductive bias) for the voxel-wise decoding and refined prediction. Obviously, our proposed global-local cube-tree fusion module focuses merely on all border vulnerable points in Fig. \ref{fig:bvp}(f) rather than all regular dense points in Fig. \ref{fig:bvp}(c), which is more related to the lung organ regions. Experimental results demonstrate the efficacy of this design.

\vspace{-0.5em}
\subsection{Network Optimization}
\vspace{-0.5em}
We define a total loss jointly optimizing the model in an end-to-end manner. The \textbf{ordinary loss} in Equ. \ref{equ:ol} is employed to supervise the first stage training of the network and produce the coarse mask predictions.
\vspace{-0.7em}
\begin{equation}\label{equ:ol}
\centering
L_{\texttt{ol}} = \sum\nolimits_{l=1}^4\{\lambda_o^l L_d(P^l,Y^l) + \lambda_o^l L_b(P^l,Y^l)\},
\vspace{-0.5em}
\end{equation}
where $L_d$, $L_b$ are Dice loss and BCE loss separately. ($P^l$,$Y^l$) is the prediction and ground truth of the segmentation in the deep layer $l$. $\lambda_o^l\in\{0.5,0.3,0.1,0.1\}$ are balance parameters. The \textbf{boundary rendering loss} in Equ. \ref{equ:brl} will supervise the training of the second stage network and produce the fine mask predictions.
\vspace{-1.0em}
\begin{equation}\label{equ:brl}
\centering
L_{\texttt{brl}} = \sum\nolimits_{l=1}^4\{\lambda_{br}^l L_d(P_{br}^l,Y_{br}^l) + \lambda_{br}^l L_b(P_{br}^l,Y_{br}^l)\},
\vspace{-0.5em}
\end{equation}
where ($P_{br}^l$,$Y_{br}^l$) is the voxel-wise border prediction and ground truth in the deep layer $l$. $\lambda_{br}^l\in\{0.5,0.3,0.1,0.1\}$ are balance parameters. The \textbf{total loss} $L = L_{\texttt{ol}} + L_{\texttt{brl}}$ consists of the ordinary loss and boundary rendering loss.

\vspace{-0.5em}
\section{Experiments}
\vspace{-0.5em}
\noindent \textbf{Datasets.} We trained and compared our model with others using chest CT scans from the public BAS airway dataset and PARSE22 \cite{luo2023efficient} artery dataset respectively. Besides, public AeroPath \cite{stoverud2023aeropath} and our in-house Lung fibrosis datasets are used for tests. BAS includes 90 cases, 20 cases from EXACT'09 and 70 cases from LIDC. (1) EXACT'09 \cite{lo2012extraction} owns 20 cases for training and 20 cases for test (without labels), scanning from normal conditions to lung disease patients. LIDC has 70 cases with labels \cite{qin2020airwaynet}. Lung fibrosis dataset has 25 labeled cases. AeroPath has 27 cases from patients with various pathologies. \textbf{Experiment setup:} We divide BAS dataset into 72/18 cases for train/test; Studies on PARSE2022 dataset follow official train/val/test split. The BAS and PARSE22 scans are both cropped as $128\times96\times144$ patches for training. All modules are trained by sample random flip for 120 epochs, an initial learning rate of $10^{-3}$, an AdamW optimizer. The whole project is realized by Pytorch \& MinkowskiEngine libraries.
\vspace{-0.5em}
\subsection{Qualitative analysis}
\vspace{-0.45em}
\begin{figure}[!htb]
\vspace{-0.5em}
\centering
\subfigure[datasets]{
\begin{minipage}[b]{0.13\linewidth}
\includegraphics[width=1\linewidth]{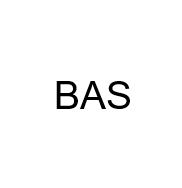}\vspace{2pt}
\includegraphics[width=1\linewidth]{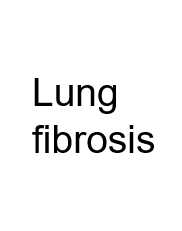}
\end{minipage}}
\subfigure[GT]{
\begin{minipage}[b]{0.14\linewidth}
\includegraphics[width=1\linewidth,height=\textwidth]{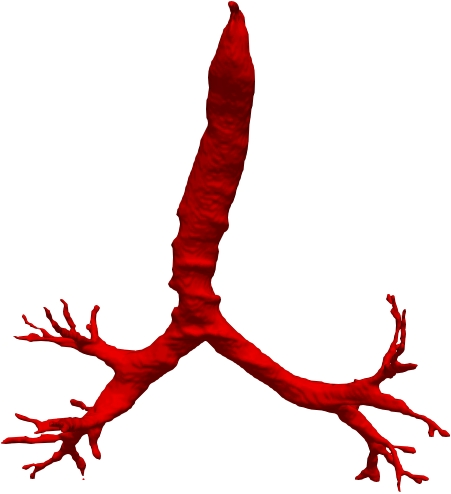}\vspace{2pt}
\includegraphics[width=1\linewidth,height=\textwidth]{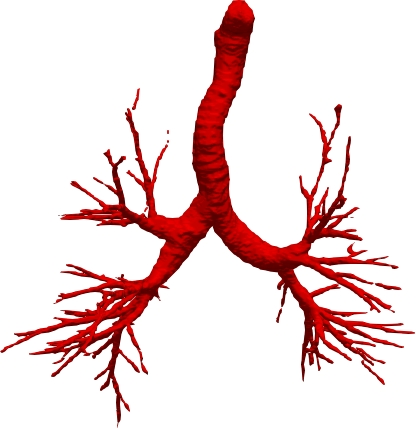}
\end{minipage}}
\subfigure[SFCN]{
\begin{minipage}[b]{0.14\linewidth}
\includegraphics[width=1\linewidth,height=\textwidth]{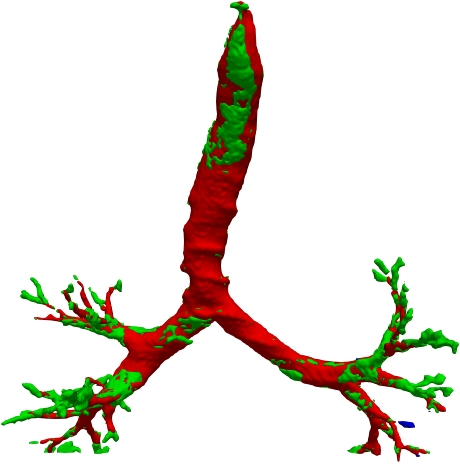}\vspace{2pt}
\includegraphics[width=1\linewidth,height=\textwidth]{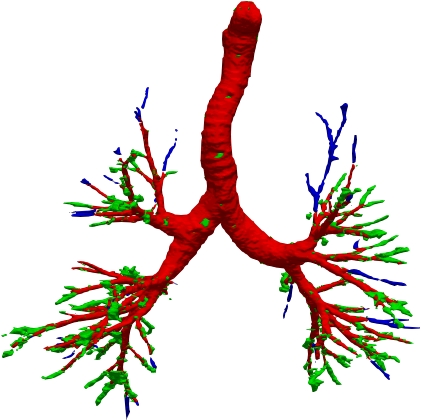}
\end{minipage}}
\subfigure[WNet]{
\begin{minipage}[b]{0.14\linewidth}
\includegraphics[width=1\linewidth,height=\textwidth]{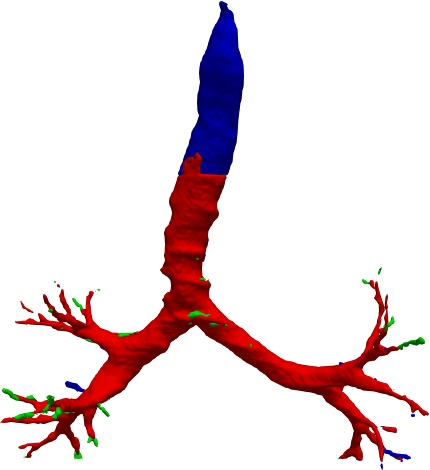}\vspace{2pt}
\includegraphics[width=1\linewidth,height=\textwidth]{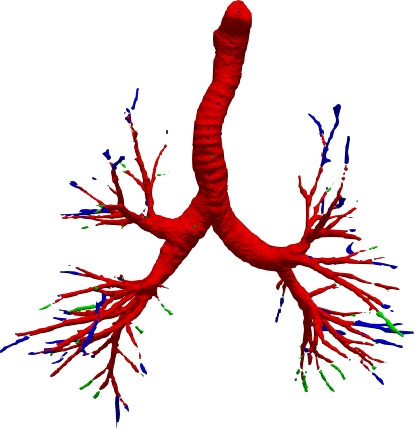}
\end{minipage}}
\subfigure[FANN]{
\begin{minipage}[b]{0.14\linewidth}
\includegraphics[width=1\linewidth,height=\textwidth]{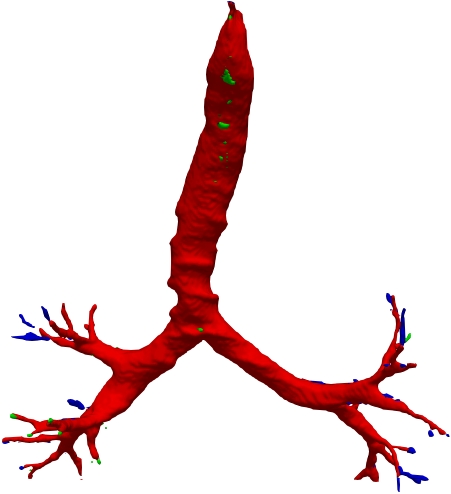}\vspace{2pt}
\includegraphics[width=1\linewidth,height=\textwidth]{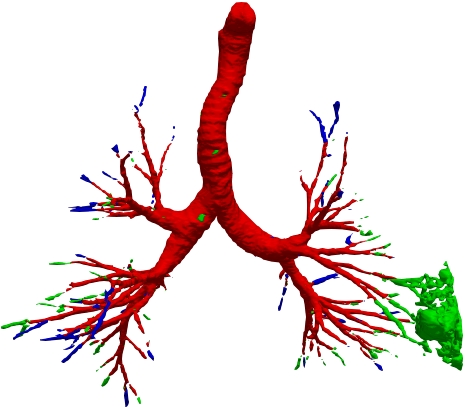}
\end{minipage}}
\subfigure[Ours]{
\begin{minipage}[b]{0.14\linewidth}
\includegraphics[width=1\linewidth,height=\textwidth]{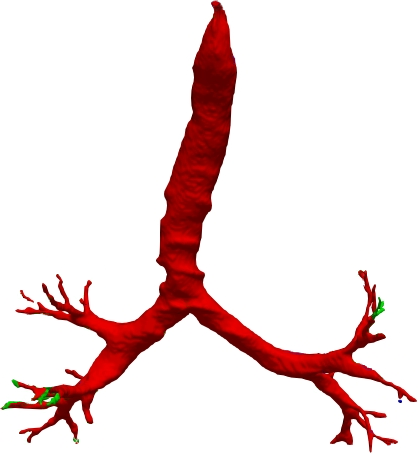}\vspace{2pt}
\includegraphics[width=1\linewidth,height=\textwidth]{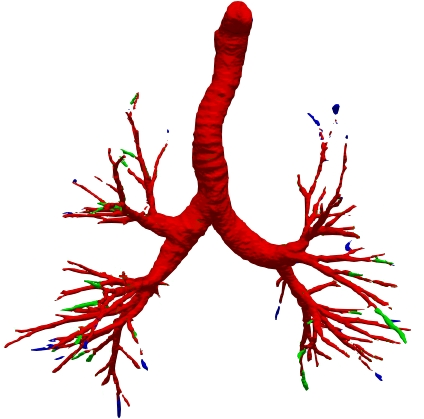}
\end{minipage}}
\vspace{-1.0em}
\caption{Qualitative airway segmentation on BAS/Lung fibrosis datasets. GT: ground truth. Red color: true positive. Green color: false positive. Blue color: false negative.}
\label{fig:qual}
\vspace{-1.5em}
\end{figure}
\noindent We qualitatively analyze our method on four challenging lung organ datasets. In Fig. 5, SFCN \cite{wang2019tubular} suffers from severe false positives and some false negatives, especially for the big green areas of airway leakages. WNet \cite{zheng2021alleviating} is mainly influenced by false negatives on the main trachea. For the Fibrosis dataset at the third row, it also encounters the false negative problem in the terminal bronchioles moderately. FANN \cite{nan2023fuzzy} bears the slight discontinuity issue of false negative in the terminal bronchioles of BAS dataset, and the severe discontinuity and airway leakage problems on the more challenging Fibrosis benchmark. Instead, due to the above two novel modules, our method can solve the defects of false negative, discontinuity, and leakages faced by past advanced methods. Besides, the results on PARSE22 artery dataset in supplementary Fig. 6 also proves this.
\vspace{-2.0em}
\subsection{Quantitative analysis}
\vspace{-0.5em}
We accurately compare our method with other advanced models in Tables \ref{tab:covid}-\ref{tab:parse22}.\\
\indent\textbf{Evaluation metrics.} The metrics are diverse, including IoU, precision, DLR, DBR, AMR, and an union metric CCFs \cite{nan2023fuzzy} that concurrently evaluates the core factors of continuity \& completeness for airway \& artery segmentation. Besides, DBR ($= N_x/N_y$) is the ratio of correctly identified branches' number $N_x$ (IoU > 0.8) to ground-truths' $N_y$. DLR ($= L_x/L_y$) is the ratio of correctly detected branch total length $L_x$ to that of ground-truths $L_y$. AMR ($= V_x/V_y$) is the ratio of false-negative volumes $V_x$ to ground-truths' $V_y$.\\ 
\begin{table}[!htb]
\vspace{-2.2em}
\centering
\scriptsize
\caption{Comparisons on the public BAS/Lung fibrosis datasets. All values are denoted by the percentage (\%) of mean/std. Red font are the best results. DLR/DBR: detected length/branch ratio, AMR: airway missing ratio. ``$\star$'' depicts statistical significance (with Wilcoxon signed-rank test p-value $<$ 0.05) compared with our method.}\label{tab:covid}
\vspace{0.5em}
\begin{tabular}{|c|c|c|c|c|c|c|}
\hline
\multirow{2}{*}{Methods} & \multicolumn{6}{c|}{\textbf{BAS}} \\
\cline{2-7}
 & IoU $\uparrow$ & Precision $\uparrow$ & DLR $\uparrow$ & DBR $\uparrow$ & AMR $\downarrow$ & CCFs $\uparrow$ \\
\hline
nnUNet \cite{isensee2021nnu} & \textcolor{red}{88.05/3.13} & \textcolor{red}{94.36/2.34$^\star$} & 86.84/7.00$^\star$ & 79.21/9.43$^\star$ & 6.96/4.02$^\star$ & 87.50/4.16$^\star$ \\
NaviAir \cite{wang2023naviairway} & 83.53/3.32$^\star$ & 86.76/4.01$^\star$ & 87.34/7.16$^\star$ & 81.01/9.52$^\star$ & \textcolor{red}{4.13/3.04$^\star$} & 85.01/3.57$^\star$  \\
PSAR \cite{tang2023adversarial} & 81.33/5.18 & 86.00/4.01 & 89.02/9.67 & 84.39/12.61 & 6.23/5.05 & --/--  \\
FANN \cite{nan2023fuzzy} & 87.38/4.45 & 91.87/3.20 & 92.71/7.93$^\star$ & 89.01/10.3$^\star$ & 5.22/4.50 & 89.69/5.54$^\star$  \\
\hline
Ours & 87.91/3.07 & 92.32/3.36 & \textcolor{red}{95.61/4.55} & \textcolor{red}{93.29/5.75} & 5.46/3.34 & \textcolor{red}{91.12/3.22}  \\
\hline
\multirow{2}{*}{Methods} & \multicolumn{6}{c|}{\textbf{Lung fibrosis}} \\
\cline{2-7}
 & IoU $\uparrow$ & Precision $\uparrow$ & DLR $\uparrow$ & DBR $\uparrow$ & AMR $\downarrow$ & CCFs $\uparrow$ \\
\hline
nnUNet \cite{isensee2021nnu} & 83.12/4.95$^\star$ & \textcolor{red}{93.81/3.14$^\star$} & 58.15/6.80$^\star$ & 50.18/7.93$^\star$ & 11.74/2.93$^\star$ & 69.72/5.64$^\star$  \\
NaviAir \cite{wang2023naviairway} & 80.79/5.33$^\star$ & 92.51/1.61$^\star$ & 59.93/14.41$^\star$ & 51.47/14.89$^\star$ & 13.45/6.45$^\star$ & 69.08/11.60$^\star$  \\
PSAR \cite{tang2023adversarial} & 72.72/6.31 & 78.79/8.16 & 72.42/10.96 & 65.50/12.66 & 9.16/3.25 & --/--  \\
FANN \cite{nan2023fuzzy} & 82.69/4.02$^\star$ & 89.04/3.73 & 78.98/8.00$^\star$ & 73.44/9.54$^\star$ & 7.95/2.37$^\star$ & 80.99/5.17$^\star$  \\
\hline
Ours & \textcolor{red}{83.81/4.64} & 89.87/4.12 & \textcolor{red}{85.10/8.58} & \textcolor{red}{80.01/10.17} & \textcolor{red}{7.10/2.33} & \textcolor{red}{84.39/5.58}  \\
\hline
\end{tabular}
\vspace{-2.5em}
\end{table}

\begin{table}[!htb]
\vspace{-1.0em}
\centering
\scriptsize
\caption{Comparison on the public validation set of PARSE22. All values are from the official evaluation with the percentage (\%) of multi-level dice coefficient.}\label{tab:parse22}
\vspace{0.5em}
\begin{tabular}{|c|c|c|c|c|c|c|c|c|c|c|c|c|}
\hline
\multirow{2}{*}{Methods} & \multicolumn{4}{c|}{\textbf{Main artery}} & \multicolumn{4}{c|}{\textbf{Branch artery}} & \multicolumn{4}{c|}{\textbf{Weighted Average}} \\
\cline{2-13}
 & 25pc $\uparrow$ & 50pc $\uparrow$ & 75pc $\uparrow$ & mean & 25pc $\uparrow$ & 50pc $\uparrow$ & 75pc $\uparrow$ & mean & 25pc $\uparrow$ & 50pc $\uparrow$ & 75pc $\uparrow$ & mean \\
\hline
NaviAir \cite{wang2023naviairway} & 84.50 & 88.63 & 89.87 & 87.11 & 55.87 & 62.85 & 66.41 & 61.40 & 63.05 & 67.77 & 70.72 & 66.54 \\
nnUNet \cite{isensee2021nnu} & 89.51 & 92.63 & \textcolor{red}{94.96} & 91.33 & \textcolor{red}{79.77} & 85.48 & \textcolor{red}{87.71} & 82.54 & 81.82 & 86.69 & \textcolor{red}{88.88} & 84.29 \\
FANN \cite{nan2023fuzzy} & 90.31 & 92.55 & 94.16 & 91.96 & 75.23 & 81.74 & 84.81 & 80.19 & 78.54 & 84.36 & 86.26 & 82.54 \\
\hline
Ours & \textcolor{red}{91.73} & \textcolor{red}{92.85} & 94.60 & \textcolor{red}{92.27} & 79.15 & \textcolor{red}{85.71} & 87.41 & \textcolor{red}{83.13} & \textcolor{red}{81.87} & \textcolor{red}{87.36} & 88.80 & \textcolor{red}{84.96} \\
\hline
\end{tabular}
\vspace{-1.5em}
\end{table}
\indent\textbf{Comparison on BAS dataset.} In the top of Table \ref{tab:covid}, our FABR obtains the best performance with a 91.12\% CCFs, 95.61\% DLR, and 93.29\% DBR. NaviAir \cite{wang2023naviairway} has the lowest AMR (4.13\%), while it performs poorly at the metrics of 83.53\% IoU, 86.76\% precision and 81.01\% DBR. Even if nnUNet \cite{isensee2021nnu} acquires the best IoU and precision scores, its DLR and DBR metrics are unsatisfied. FANN 
achieves a suboptimal performance (89.69\% CCFs, 92.71\% DLR, 89.01\%DBR).\\
\indent\textbf{Comparison on fibrosis dataset.} Although it's the very challenging benchmark, our FABR still behaves robustly and exceeds the best method FANN by 3.4\% CCFs with a total metrics of 84.39\% CCFs, 83.81\% IoU, 85.1\% DLR, 80.01\% DBR. The lowest AMR (7.1\%) confirms that our method can solve the discontinuity issue well. Other methods also behave similarly to the BAS dataset. As seen in the two datasets, the main improvements of our method are consistently at the IoU, DLR and DBR metrics, which are mainly influenced by bronchioles and trachea borders that are easily lost due to network down/up-samples. Hence, our method can extract the robust bronchiole features and render border well via the two novel modules for the accurate lung organ segmentation.\\
\indent\textbf{Comparison on PARSE22 dataset.} This dataset is more challenging due to more dense small bronchioles shown in supplementary Fig. 6. However, our method still reaches the best weighted average multi-level dice of 84.96\% in Table \ref{tab:parse22} compared against some advanced methods via the official evaluation. As you can see, the remarkable gain comes from the ``branch artery'', which maintains the consistency with above airway segmentation.\\
\indent\textbf{Ablation studies.} To verify the efficacy of each module, we perform the thorough ablation studies in supplementary Tables 3-5 and Figs. 7-8. In Table 3, the 2-\textit{th} row on lung fibrosis dataset with the proposed FA-based transformer-like backbone achieves the largest 2.24\% $\triangle$CCFs, verifying the efficacy of fusing fuzzy sets and deep network to diminish the uncertainty in feature representations significantly. The 3-\textit{th} row with GLCF module indicates 1.02\% $\triangle$CCFs, proving that we only need to focus much on the very hard BVP rather than all regular dense points, which provide the most important losing information of discontinuity or details in the network down-sample operation. Since we only extract the BVP to render, it can suppress the redundant background to further solve the severe class imbalance issue of foreground and background voxels. Supplementary Table 4 evidences the efficacy of GLCF module which improves the border accuracy obviously by 4.72\%. In Table 5, the 2-\textit{th} row with FA-based transformer-like backbone improves the DBR significantly on the terminal (1.8\%), small (1.25\%) and medium (1.65\%) branches except the large trachea (-1.03\%), for most uncertainty in the feature representations is from the terminal, small and medium branches that are too thin and hard to be discerned while annotating. The 3-\textit{th} row with GLCF module realizes the significant promotion of DBR on the small (2.02\%), medium (2.02\%) and large (3.09\%) branches, which is consistent with Fig. 8 to overcome the issue of detail loss in the network down-sample operation and render the BVP effectively. Supplementary Fig. 7 elucidates that our FA-based transformer-like backbone can enhance the feature representations of lung organs significantly.

\vspace{-1.2em}
\section{Conclusion}
\vspace{-0.5em}
Automated lung organ segmentation is vital to aid radiologists with lung disease diagnosis and prognosis. However, most prior top-tier methods suffer from the discontinuity, false-negative and leakage issues. Inspired by these, we proposed the innovative method FABR in the paper, which has two novel modules, i.e., (1) Fuzzy attention-based transformer-like backbone, diminishing the uncertainty of lung organ feature representations; (2) The global-local cube-tree feature fusion module, explicitly modeling the border vulnerable points yielded by recycled down/up-sample for accurate lung organ segmentation. Finally, extensive qualitative and quantitative experiments have proven the excellent performance of our method on four challenging lung organ segmentation datasets, involving CT scans of lung cancer, fibrosis, and mild lung diseases.

\begin{credits}
\subsubsection{\ackname}
\vspace{-0.5em}
The study was supported in part by ERC IMI (101005122), H2020 (952172), MRC (MC/PC/21013), the Royal Society (IEC/NSFC/211235), NVIDIA Academic Hardware Grant Program, SABER project funded by Boehringer Ingelheim Ltd, NIHR Imperial Biomedical Research Centre (RDA01), Wellcome Leap Dynamic Resilience, UKRI Future Leaders Fellowship (MR/V023799/1), and UKRI Fellowship (EP/Z002206/1).
\vspace{-0.2em}
\subsubsection{\discintname}
\vspace{-0.5em}
The authors have no competing interests to declare that are relevant to the content of this article.
\vspace{-0.5em}
\end{credits}

%
%
%
\bibliographystyle{splncs04}
\bibliography{miccai}

\begin{thebibliography}{10}
\providecommand{\url}[1]{\texttt{#1}}
\providecommand{\urlprefix}{URL }
\providecommand{\doi}[1]{https://doi.org/#1}

\bibitem{dosovitskiy2020image}
Dosovitskiy, A., Beyer, L., Kolesnikov, A., Weissenborn, D., Zhai, X.,
  Unterthiner, T., Dehghani, M., Minderer, M., Heigold, G., Gelly, S., et~al.:
  An image is worth 16x16 words: Transformers for image recognition at scale.
  arXiv preprint arXiv:2010.11929  (2020)

\bibitem{fang2024dynamic}
Fang, Y., Wu, S., Zhang, S., Huang, C., Zeng, T., Xing, X., Walsh, S., Yang,
  G.: Dynamic multimodal information bottleneck for multimodality
  classification. In: Proceedings of the IEEE/CVF Winter Conference on
  Applications of Computer Vision. pp. 7696--7706 (2024)

\bibitem{gao2021future}
Gao, X., Jin, Y., Zhao, Z., Dou, Q., Heng, P.A.: Future frame prediction for
  robot-assisted surgery. In: Information Processing in Medical Imaging: 27th
  International Conference, IPMI 2021, Virtual Event, June 28--June 30, 2021,
  Proceedings 27. pp. 533--544. Springer (2021)

\bibitem{hatamizadeh2022unetr}
Hatamizadeh, A., Tang, Y., Nath, V., Yang, D., Myronenko, A., Landman, B.,
  Roth, H.R., Xu, D.: Unetr: Transformers for 3d medical image segmentation.
  In: Proceedings of the IEEE/CVF winter conference on applications of computer
  vision. pp. 574--584 (2022)

\bibitem{hu2020squeeze}
Hu, J., Shen, L., Albanie, S., Sun, G.: Squeeze-and-excitation networks. IEEE
  transactions on pattern analysis and machine intelligence  \textbf{42}(8),
  2011--2023 (2020)

\bibitem{isensee2021nnu}
Isensee, F., Jaeger, P.F., Kohl, S.A., Petersen, J., Maier-Hein, K.H.: nnu-net:
  a self-configuring method for deep learning-based biomedical image
  segmentation. Nature methods  \textbf{18}(2),  203--211 (2021)

\bibitem{lin2021seg4reg+}
Lin, Y., Liu, L., Ma, K., Zheng, Y.: Seg4reg+: Consistency learning between
  spine segmentation and cobb angle regression. In: Medical Image Computing and
  Computer Assisted Intervention--MICCAI 2021: 24th International Conference,
  Strasbourg, France, September 27--October 1, 2021, Proceedings, Part V 24.
  pp. 490--499. Springer (2021)

\bibitem{lo2012extraction}
Lo, P., Van~Ginneken, B., Reinhardt, J.M., Yavarna, T., De~Jong, P.A., Irving,
  B., Fetita, C., Ortner, M., Pinho, R., Sijbers, J., et~al.: Extraction of
  airways from ct (exact'09). IEEE Transactions on Medical Imaging
  \textbf{31}(11),  2093--2107 (2012)

\bibitem{luo2023efficient}
Luo, G., Wang, K., Liu, J., Li, S., Liang, X., Li, X., Gan, S., Wang, W., Dong,
  S., Wang, W., et~al.: Efficient automatic segmentation for multi-level
  pulmonary arteries: The parse challenge. arXiv preprint arXiv:2304.03708
  (2023)

\bibitem{nan2023fuzzy}
Nan, Y., Del~Ser, J., Tang, Z., Tang, P., Xing, X., Herrera, F., Pedrycz, W.,
  Walsh, S., Yang, G.: Fuzzy attention neural network to tackle discontinuity
  in airway segmentation. IEEE Transactions on Neural Networks and Learning
  Systems  (2023)

\bibitem{nan2024hunting}
Nan, Y., Xing, X., Wang, S., Tang, Z., Felder, F.N., Zhang, S., Ledda, R.E.,
  Ding, X., Yu, R., Liu, W., et~al.: Hunting imaging biomarkers in pulmonary
  fibrosis: Benchmarks of the aiib23 challenge. Medical Image Analysis p.
  103253 (2024)

\bibitem{oktay2018attention}
Oktay, O., Schlemper, J., Folgoc, L.L., Lee, M., Heinrich, M., Misawa, K.,
  Mori, K., McDonagh, S., Hammerla, N.Y., Kainz, B., et~al.: Attention u-net:
  Learning where to look for the pancreas. arXiv preprint arXiv:1804.03999
  (2018)

\bibitem{qin2020airwaynet}
Qin, Y., Gu, Y., Zheng, H., Chen, M., Yang, J., Zhu, Y.M.: Airwaynet-se: A
  simple-yet-effective approach to improve airway segmentation using context
  scale fusion. In: 2020 IEEE 17th International Symposium on Biomedical
  Imaging (ISBI). pp. 809--813. IEEE (2020)

\bibitem{ronneberger2015u}
Ronneberger, O., Fischer, P., Brox, T.: U-net: Convolutional networks for
  biomedical image segmentation. In: Medical Image Computing and
  Computer-Assisted Intervention--MICCAI 2015: 18th International Conference,
  Munich, Germany, October 5-9, 2015, Proceedings, Part III 18. pp. 234--241.
  Springer (2015)

\bibitem{stoverud2023aeropath}
St{\o}verud, K.H., Bouget, D., Pedersen, A., Leira, H.O., Lang{\o}, T.,
  Hofstad, E.F.: Aeropath: An airway segmentation benchmark dataset with
  challenging pathology. arXiv preprint arXiv:2311.01138  (2023)

\bibitem{tang2023adversarial}
Tang, Z., Nan, Y., Walsh, S., Yang, G.: Adversarial transformer for repairing
  human airway segmentation. IEEE Journal of Biomedical and Health Informatics
  (2023)

\bibitem{tsay2021lower}
Tsay, J.C.J., Wu, B.G., Sulaiman, I., Gershner, K., Schluger, R., Li, Y., Yie,
  T.A., Meyn, P., Olsen, E., Perez, L., et~al.: Lower airway dysbiosis affects
  lung cancer progression. Cancer discovery  \textbf{11}(2),  293--307 (2021)

\bibitem{wang2023naviairway}
Wang, A., Tam, T.C.C., Poon, H.M., Yu, K.C., Lee, W.N.: Naviairway: a
  bronchiole-sensitive deep learning-based airway segmentation pipeline for
  planning of navigation bronchoscopy. Authorea Preprints  (2023)

\bibitem{wang2019tubular}
Wang, C., Hayashi, Y., Oda, M., Itoh, H., Kitasaka, T., Frangi, A.F., Mori, K.:
  Tubular structure segmentation using spatial fully connected network with
  radial distance loss for 3d medical images. In: Medical Image Computing and
  Computer Assisted Intervention--MICCAI 2019: 22nd International Conference,
  Shenzhen, China, October 13--17, 2019, Proceedings, Part VI 22. pp. 348--356.
  Springer (2019)

\bibitem{woo2023convnext}
Woo, S., Debnath, S., Hu, R., Chen, X., Liu, Z., Kweon, I.S., Xie, S.: Convnext
  v2: Co-designing and scaling convnets with masked autoencoders. In:
  Proceedings of the IEEE/CVF Conference on Computer Vision and Pattern
  Recognition. pp. 16133--16142 (2023)

\bibitem{yang2022st++}
Yang, L., Zhuo, W., Qi, L., Shi, Y., Gao, Y.: St++: Make self-training work
  better for semi-supervised semantic segmentation. In: Proceedings of the
  IEEE/CVF Conference on Computer Vision and Pattern Recognition. pp.
  4268--4277 (2022)

\bibitem{zheng2021alleviating}
Zheng, H., Qin, Y., Gu, Y., Xie, F., Yang, J., Sun, J., Yang, G.Z.: Alleviating
  class-wise gradient imbalance for pulmonary airway segmentation. IEEE
  transactions on medical imaging  \textbf{40}(9),  2452--2462 (2021)

\end{thebibliography}

\appendix
\section{Supplementary}
\vspace{-1.5em}
\begin{figure}[!htb]
\setcounter{figure}{5}
\centering
\subfigure[Lung artery slice]{
\begin{minipage}[b]{0.3\linewidth}
\centering
\includegraphics[width=0.75\linewidth,height=0.675\linewidth]{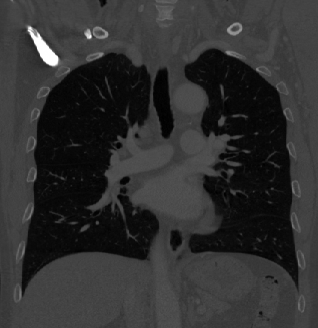}
\end{minipage}}
\subfigure[GT]{
\begin{minipage}[b]{0.3\linewidth}
\centering
\includegraphics[width=1\linewidth,height=0.675\linewidth]{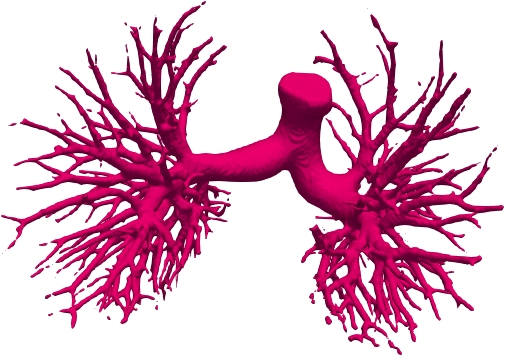}
\end{minipage}}
\subfigure[WNet]{
\begin{minipage}[b]{0.3\linewidth}
\centering
\includegraphics[width=1\linewidth,height=0.675\linewidth]{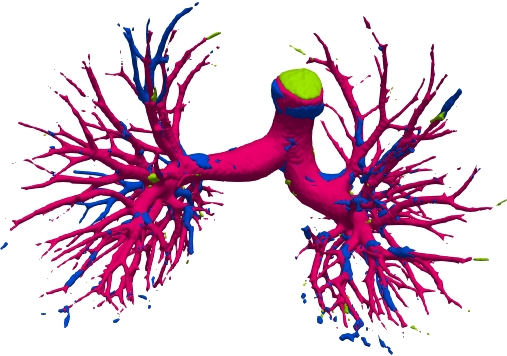}
\end{minipage}}
\subfigure[nnUNet]{
\begin{minipage}[b]{0.3\linewidth}
\centering
\includegraphics[width=1\linewidth,height=0.675\linewidth]{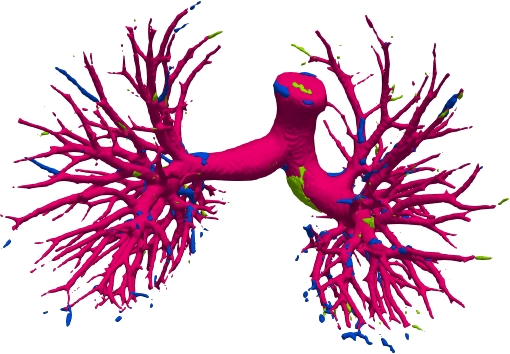}
\end{minipage}}
\subfigure[FANN]{
\begin{minipage}[b]{0.3\linewidth}
\centering
\includegraphics[width=1\linewidth,height=0.675\linewidth]{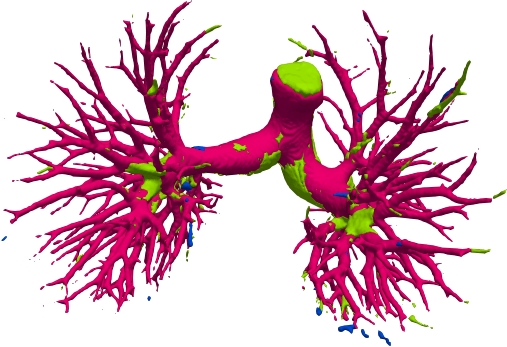}
\end{minipage}}
\subfigure[Ours]{
\begin{minipage}[b]{0.3\linewidth}
\centering
\includegraphics[width=1\linewidth,height=0.675\linewidth]{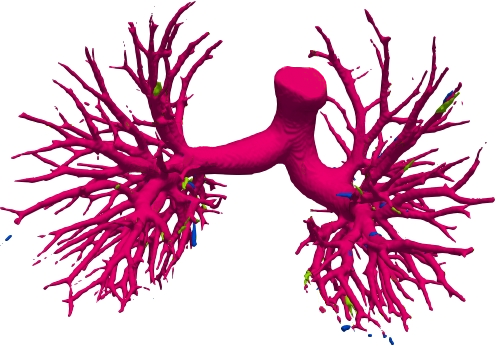}
\end{minipage}}
\vspace{-0.7em}
\caption{Comparison with other state-of-the-art methods on PARSE22 lung artery dataset. Light red color: true positive. Light green color: false positive. Light blue color: false negative.}\label{fig:parse}
\vspace{-3.5em}
\end{figure}

\begin{figure}[!htb]
\centering
\subfigure[Lung airway slice]{
\begin{minipage}[b]{0.25\linewidth}
\centering
\includegraphics[width=\linewidth,height=0.675\linewidth]{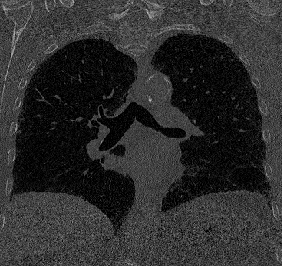}
\end{minipage}}
\hspace{1em}
\subfigure[Airway without our FA-backbone]{
\begin{minipage}[b]{0.25\linewidth}
\centering
\includegraphics[width=1\linewidth,height=0.675\linewidth]{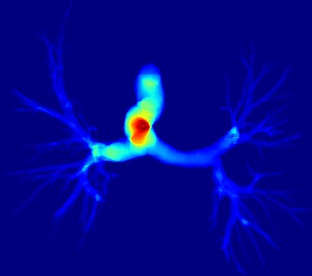}
\end{minipage}}
\subfigure[Airway with our FA-backbone]{
\begin{minipage}[b]{0.4\linewidth}
\centering
\includegraphics[width=0.625\linewidth,height=0.422\linewidth]{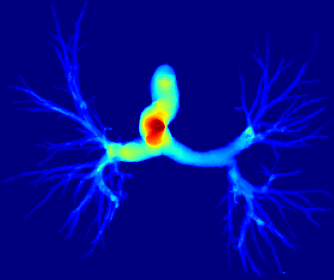}
\includegraphics[width=0.12\linewidth,height=0.422\linewidth]{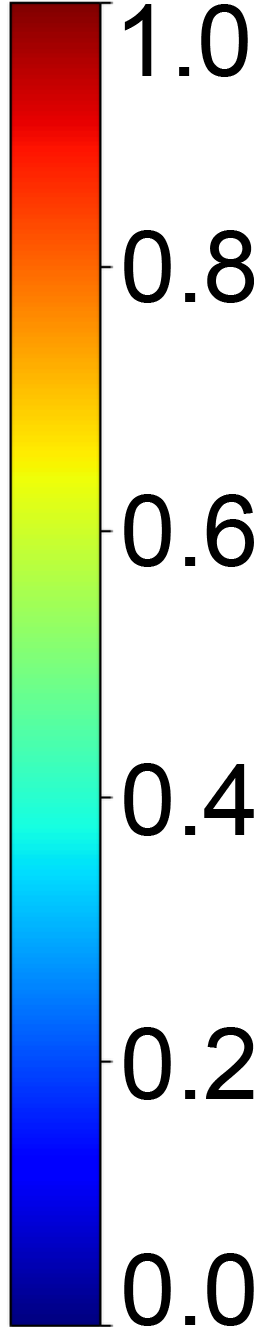}
\end{minipage}}
\subfigure[Lung artery slice]{
\begin{minipage}[b]{0.25\linewidth}
\centering
\includegraphics[width=1\linewidth,height=0.675\linewidth]{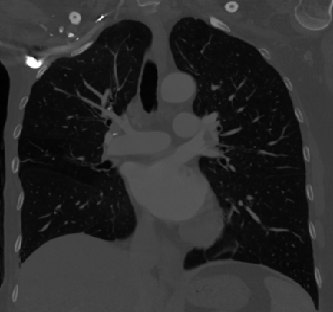}
\end{minipage}}
\hspace{1.em}
\subfigure[Artery without our FA-backbone]{
\begin{minipage}[b]{0.25\linewidth}
\centering
\includegraphics[width=1\linewidth,height=0.675\linewidth]{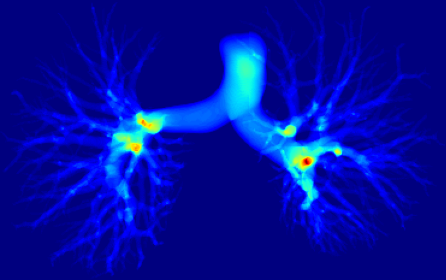}
\end{minipage}}
\subfigure[Artery with our FA-backbone]{
\begin{minipage}[b]{0.4\linewidth}
\centering
\includegraphics[width=0.625\linewidth,height=0.422\linewidth]{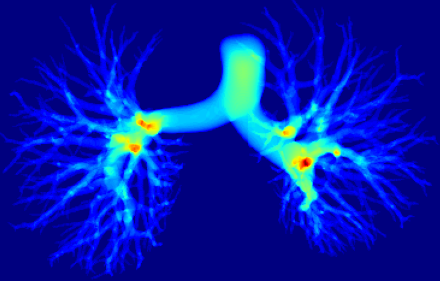}
\includegraphics[width=0.12\linewidth,height=0.422\linewidth]{figs/94_01_white2.png}
\end{minipage}}
\caption{Visualization with the proposed fuzzy attention-based transformer-like backbone or not on lung organ datasets.}\label{fig:fab}
\vspace{-0.5em}
\end{figure}

\begin{figure}[!htb]
\centering
\subfigure[Lung artery slice]{
\begin{minipage}[b]{0.3\linewidth}
\centering
\includegraphics[width=1\linewidth,height=0.675\linewidth]{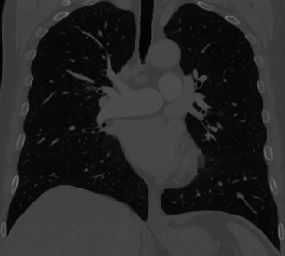}
\end{minipage}}
\hspace{2mm}
\subfigure[Without border rendering]{
\begin{minipage}[b]{0.3\linewidth}
\centering
\includegraphics[width=1\linewidth,height=0.675\linewidth]{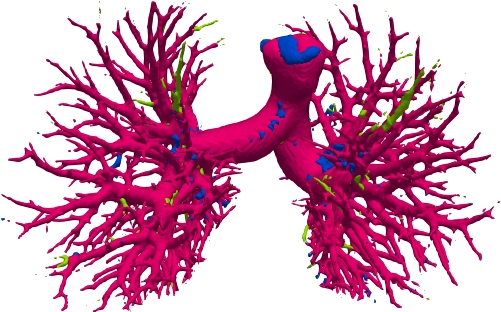}
\end{minipage}}
\hspace{2mm}
\subfigure[With border rendering]{
\begin{minipage}[b]{0.3\linewidth}
\centering
\includegraphics[width=1\linewidth,height=0.675\linewidth]{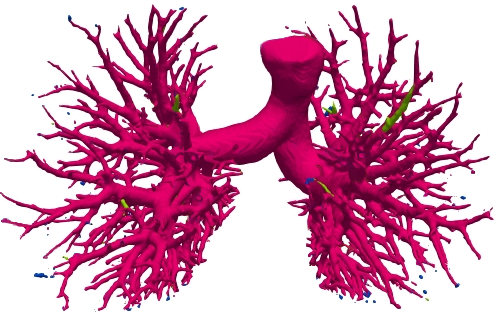}
\end{minipage}}
\caption{Comparison with border rendering modules or not on PARSE22 dataset. Light red color: true positive. Light green color: false positive. Light blue color: false negative.}\label{fig:aba_br}
\vspace{-1.em}
\end{figure}

\begin{table}[!htb]
\setcounter{table}{2}
\centering
\scriptsize
\caption{Ablation studies of the proposed modules on Lung fibrosis dataset. BL: reproduced FANN. FA: fuzzy attention-based transformer-like backbone. GLCF: \textbf{g}lobal-\textbf{l}ocal \textbf{c}ube-tree \textbf{f}usion. $\triangle$CCFs$\uparrow$: the difference of adjacent two rows.}\label{tab:aba}
\vspace{1.0em}
\begin{tabular}{|c|c|c|c|c|c|c|c|c|c|}
\hline
\multicolumn{3}{|c|}{Proposed modules} & \multicolumn{7}{c|}{\textbf{Lung fibrosis}} \\
\hline
\ \ BL\ \ &\ \ FA\ \ & GLCF & IoU $\uparrow$ & Precision $\uparrow$ & DLR $\uparrow$ & DBR $\uparrow$ & AMR $\downarrow$ & CCFs $\uparrow$ & $\triangle$CCFs $\uparrow$ \\
\hline
$\surd$ & -- & -- & 82.52/4.36 & 88.90/3.82 & 79.41/8.22 & 73.90/9.85 & 7.79/2.46 & 81.13/5.34 & --- \\
$\surd$ & $\surd$ & -- & 83.35/4.11 & 89.29/3.62 & 83.44/7.86 & 78.02/9.41 & 7.32/2.73 & 83.37/4.42 & \textcolor{red}{2.24/-0.92} \\
$\surd$ & $\surd$ & $\surd$ & 83.81/4.64 & 89.87/4.12 & 85.10/8.58 & 80.01/10.17 & 7.10/2.33 & 84.39/5.58 & \textcolor{blue}{1.02/1.16} \\
\hline
\end{tabular}
\vspace{-1.2em}
\end{table}

\begin{table}[!ht]
\centering
\scriptsize
\caption{The efficacy of proposed module on border rendering on the pathological dataset AeroPath.
$\triangle$: the difference of adjacent two rows.}\label{tab:aba_bo}
\vspace{1.0em}
\begin{tabular}{|c|c|c|c|}
\hline
without GLCF & with GLCF & Accuracy & $\triangle$Accuracy \\
\hline
$\surd$ & -- & 71.63/16.12 & -- \\
-- & $\surd$ & 76.35/16.46 & 4.72/0.34 \\
\hline
\end{tabular}
\end{table}

\begin{figure}[!ht]
  \begin{minipage}[b]{0.3\linewidth}
    \centering
    \scriptsize
    \includegraphics[width = 0.85\linewidth]{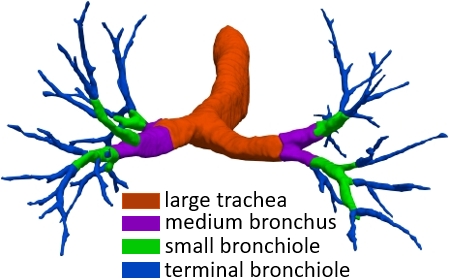}
    \caption{Varied branches.}\label{fig:br_size}
  \end{minipage}\hspace{0.5em}
  \begin{minipage}[b]{0.6\linewidth}
    \centering
    \scriptsize
    \begin{tabular}{|c|c|c|c|c|c|c|c|}
    \hline
    \multicolumn{3}{|c|}{Proposed modules} & \multicolumn{5}{c|}{\textbf{Branch sizes}} \\
    \hline
   \ \ BL\ \ &\ \ FA\ \ & GLCF & Terminal & Small & Medium & Trachea & Average \\
    \hline
   $\surd$ & -- & -- & 86.63 & 91.78 & 91.98 & 95.77 & 89.11 \\
   $\surd$ & $\surd$ & -- & 88.43 & 93.03 & 93.63 & 94.74 & 90.70  \\
   $\surd$ & $\surd$ & $\surd$ & 89.11 & 95.05 & 95.65 & 97.83 &  91.67 \\
  \hline
  \end{tabular}
  \captionof{table}{The DBR without standard deviation for varied branches on AeroPath dataset.}\label{tab:br_size}
  \end{minipage}
\end{figure}

\end{document}